# Loop Current and Antiferromagnetic States in Fermionic Hubbard Model with Staggered Flux at Half Filling

Yuta Toga[a,*] and Hisatoshi Yokoyama[b]

[a]*ESICMM, National Institute for Materials Science, Tsukuba 305-0047, Japan*
[b]*Department of Physics, Tohoku University, Sendai 980-8578, Japan*

**Abstract**

Anticipating realization of interacting fermions in an optical lattice with a large gauge field, we consider phase transitions and loop currents in a two-dimensional $S = 1/2$ fermionic-Hubbard model with $\pi/2$-staggered flux at half filling. We use a variational Monte Carlo method, which is reliable even for strong correlations. As a trial wave function, a coexistent state of antifferromagnetic and staggered-flux orders is studied. In a strongly correlated regime, the ground state becomes an insulating coexistent state with loop currents. By comparing fermions with bosons, we discuss an important role of Pauli principle.

## 1. Introductions

 Cold atoms in optical lattices provide an opportunity for studying strongly correlated systems in extremely clean and well-controlled environment [1]. The condition of charge neutrality for cold atoms, however, prevents us from studying phenomena related to charged particles in a magnetic flux. Therefore, instead, many experimentalists have made efforts to realize artificial magnetic fluxes [2]. Among them, Aidelsburger et al. proposed a useful way to realize both staggered and uniform strong artificial magnetic fluxes, using laser-assisted tunneling on bosonic lattices [3]. Afterward, properties in artificial magnetic fluxes came to attract attention as intriguing research subjects of strongly correlated lattice bosons. Similarly, fermionic counterparts are also desired to be experimentally realized.

 With such experimental development, recently, theoretical studies on current states have been actively carried out for bosonic systems with staggered fluxes [4-7]. On the other hand, as for fermionic counterparts, studies hitherto were limited to the state with $\pi$-flux per plaquette (see Fig. 1), which is a special case with no current [8-10]. In this connection, a loop-current state in the fluxless case, namely, ordinary Hubbard model, has been studied as a possible pseudogap state in cuprates superconductors, in which time-reversal and other symmetries are broken [11].

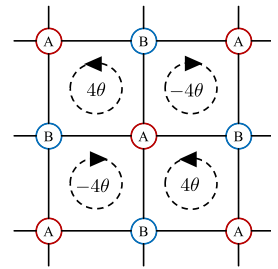

Fig.1. Schematic figure of staggered flux ($\pm 4\theta$) in square plaquettes in FHM. A and B denotes indices of the two sublattices.

 In this article, we focus on a fermionic Hubbard model with $\pi/2$-staggered fluxes per plaquette on the square lattice at half filling, and study the relationship between staggered flux (ST) and antifferromagnetic (AF) orders, using a variational Monte Carlo (VMC) method, which is useful to treat strong correlation. We compare the properties of four trial states with or without ST and AF orders. It is found that phase transitions and crossovers occur in these states, and that in a wide range of correlation strength, a coexistent state of SF and AF orders becomes stable.

---

* Corresponding author. Tel.: +81- 29-851-3354 (6704); *E-mail address:* toga.yuta@nims.go.jp

## 2. Model and Method

As a model of cold atoms in an optical lattice with magnetic fluxes, we consider an $S = 1/2$ fermionic Hubbard model (FHM) with a staggered field:

$$\mathcal{H} = -t \sum_{i \in A, \sigma} \left[ e^{i\theta} \left( \hat{a}_{i,\sigma}^\dagger \hat{b}_{i+x,\sigma} + \hat{a}_{i,\sigma}^\dagger \hat{b}_{i-x,\sigma} \right) + e^{-i\theta} \left( \hat{a}_{i,\sigma}^\dagger \hat{b}_{i+y,\sigma} + \hat{a}_{i,\sigma}^\dagger \hat{b}_{i-y,\sigma} \right) + \text{H.c.} \right] + U \sum_i \hat{n}_{i,\uparrow} \hat{n}_{i,\downarrow}, \quad (1)$$

where $a_{i,\sigma}^\dagger, b_{i,\sigma}^\dagger$ ($a_{i,\sigma}, b_{i,\sigma}$) creates (annihilates) a fermion with spin $\sigma$ at $i$ site on A and B sublattices, respectively, (see Fig. 1), $\hat{n}_{i,\sigma} = \hat{a}_{i,\sigma}^\dagger \hat{a}_{i,\sigma} = \hat{b}_{i,\sigma}^\dagger \hat{b}_{i,\sigma}$, $x$ and $y$ indicate the lattice vectors in $x$ and $y$ directions, respectively; $U$ is the on-site Hubbard repulsion, $t$ the tunneling rate, and $\theta$ the Peierls phase corresponding to a local magnetic flux. In this article, we consider a case of $4\theta = \pi/2$ (half-$\pi$ flux) at half filling.

In applying a VMC method to Eq. (1), we use trial wave functions of Jastrow type: $|\Psi\rangle = \hat{\mathcal{P}}|\Phi\rangle$, where $\hat{\mathcal{P}}$ denotes a correlation factor mentioned later and $|\Phi\rangle$ is a one-body (Hartree-Fock) part. In this work, we determine $|\Phi\rangle$ as follows: In the fluxless case ($\theta = 0$), the ground state at half filling exhibits an AF order for $U > 0$. We checked that this AF order survives as $\theta$ increases. On the other hand, for a finite $\theta$ with $U/t \to 0$, the ground state has a ST order. Hence, as $|\Phi\rangle$, a mixed state of AF and ST orders, $|\Phi_{\text{AF+ST}}\rangle$, is probably appropriate. $|\Phi_{\text{AF+ST}}\rangle$ is given as a Hartree-Fock-Bogoliubov-type wave function, and is derived from an AF mean field Hamiltonian:

$$\mathcal{H}_{\text{MF}} = \tilde{\Delta}_{\text{AF}} \sum_{\sigma, \mathbf{k} \in \text{m.b.z.}} p_\sigma \left( -\hat{a}_{\mathbf{k},\sigma}^\dagger \hat{a}_{\mathbf{k},\sigma} + \hat{b}_{\mathbf{k},\sigma}^\dagger \hat{b}_{\mathbf{k},\sigma} \right) - 2t \sum_{\sigma, \mathbf{k} \in \text{m.b.z.}} \left( u_{\mathbf{k}} \hat{a}_{\mathbf{k},\sigma}^\dagger \hat{b}_{\mathbf{k},\sigma} + u_{\mathbf{k}}^* \hat{b}_{\mathbf{k},\sigma}^\dagger \hat{a}_{\mathbf{k},\sigma} \right), \quad (2)$$

where $p_\sigma = 1$ or $-1$ according to $\sigma = \uparrow$ or $\downarrow$, m.b.z denotes the folded AF Brillouin zone, and $u_{\mathbf{k}} = \exp(i\tilde{\theta}) \cos k_x - \exp(-i\tilde{\theta}) \cos k_y$. Here $\tilde{\Delta}_{\text{AF}}$ and $\tilde{\theta}$ are variational parameters characteristic of AF and ST orders, respectively. We diagonalize $\mathcal{H}_{\text{MF}}$ by a Bogoliubov transformation to yield a single-particle band dispersion: $\epsilon_{\mathbf{k},\pm}^{\text{AF+ST}} = \pm \left( \tilde{\Delta}_{\text{AF}}^2 + 4t^2 |u_{\mathbf{k}}|^2 \right)^{1/2}$. By filling the lower band $\epsilon_{\mathbf{k},-}^{\text{AF+ST}}$, we have

$$|\Phi_{\text{AF+ST}}(\tilde{\Delta}_{\text{AF}}, \tilde{\theta})\rangle = \prod_{\sigma, \mathbf{k} \in \text{m.b.z.}} \left( \alpha_{\sigma, \mathbf{k}}^\dagger \right)^N |0\rangle, \quad (3)$$

$$\alpha_{\sigma, \mathbf{k}}^\dagger = \frac{1}{\sqrt{N_s}} \left[ \sum_{i \in A} \frac{u_{\mathbf{k}}}{|u_{\mathbf{k}}|} \sqrt{1 - \frac{\Delta_{\text{AF}} p_\sigma}{\epsilon_{\mathbf{k},-}^{\text{AF+ST}}}} e^{i\mathbf{k} \cdot \mathbf{r}_i} \hat{a}_{i,\sigma}^\dagger + \sum_{i \in B} \sqrt{1 + \frac{\Delta_{\text{AF}} p_\sigma}{\epsilon_{\mathbf{k},-}^{\text{AF+ST}}}} e^{-i\mathbf{k} \cdot \mathbf{r}_i} \hat{b}_{i,\sigma}^\dagger \right], \quad (4)$$

where $N$ ($N_s$) is the total number of fermions (sites) and $\mathbf{r}_i$ is the $i$-th site's position vector.

Now, we turn to the correlation factor $\hat{\mathcal{P}}$, which is given in the present case as,

$$\hat{\mathcal{P}} = \hat{\mathcal{P}}_\phi(\phi) \hat{\mathcal{P}}_{\text{DH}}(\eta_D, \eta_H) \hat{\mathcal{P}}_J(v_r) \hat{\mathcal{P}}_G(g), \quad (5)$$

where $\hat{\mathcal{P}}_G(g)$ and $\hat{\mathcal{P}}_J(v_r) = \exp\left[ -(1/2) \sum_{i \neq j} v(|\mathbf{r}_i - \mathbf{r}_j|) (\hat{n}_i - 1)(\hat{n}_j - 1) \right]$ are the onsite (Gutzwiller) and intersite (Jastrow) correlation projections, and $\hat{\mathcal{P}}_{\text{DH}}(\eta_D, \eta_H)$ is a projection of binding a doubly occupied site (D) and an empty (H) site in nearest-neighbor sites; $\hat{\mathcal{P}}_{\text{DH}}$ is essential for treating Mott physics. In addition, for a current-carrying state in a Mott regime, it is crucial to introduce a configuration-dependent phase factor $\hat{\mathcal{P}}_\phi(\phi)$ [11,12]. The role of $\hat{\mathcal{P}}_\phi(\phi)$ is to cancel out a Peierls phase attached in hopping processes in strongly correlated regime, where hopping is almost restricted to the case of creating or annihilating a D-H pair.

In the following, we compare four cases of $|\Psi_{\text{AF+ST}}\rangle$: $|\Psi_{\text{AF+ST}}\rangle = \hat{\mathcal{P}}|\Phi_{\text{AF+ST}}(\tilde{\Delta}_{\text{AF}}, \tilde{\theta})\rangle$, $|\Psi_{\text{ST}}\rangle = |\Psi_{\text{AF+ST}}(\tilde{\Delta}_{\text{AF}} = 0)\rangle$, $|\Psi_{\text{AF}}\rangle = |\Psi_{\text{AF+ST}}(\tilde{\theta} = 0)\rangle$, $|\Psi_{\text{FS}}\rangle = |\Psi_{\text{AF+ST}}(\tilde{\Delta}_{\text{AF}} = 0, \tilde{\theta} = 0)\rangle$. Here, $|\Psi_{\text{FS}}\rangle$ and $|\Psi_{\text{AF}}\rangle$ corresponds to the paramagnetic (Fermi sea) and AF states in the fluxless FHM ($\theta = 0$), respectively; $|\Psi_{\text{ST}}\rangle$ and $|\Psi_{\text{AF+ST}}\rangle$ can be regarded as a paramagnetic and the AF states of the FHM with staggered flux ($\theta \neq 0$), respectively.

The variational parameters ($\tilde{\Delta}_{\text{AF}}, \tilde{\theta}, g, v_r, \eta_D, \eta_H, \phi$) are optimized numerically, by using the stochastic reconfiguration method [13,14] for each set of model parameters ($U, L$), and then calculate physical quantities with 1-2$\times 10^6$ samples for $L \times L$-site lattices ($L = 12, 14$) under the periodic boundary conditions.

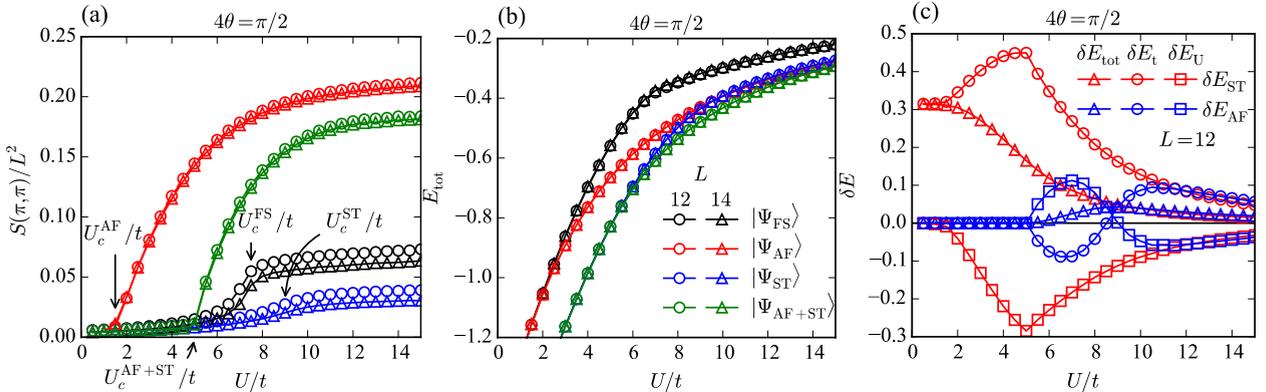

Fig. 2. (a) Spin structure factor $S(\mathbf{q})$ at $\mathbf{q}=(\pi,\pi)$ is compared among the four wave functions treated here as a function of $U/t$ in the $\pi/2$-flux case for two system sizes ($L=12,14$). (b) Total energy is similarly compared. (c) Energy differences, $\delta E_{ST} = E_{AF+ST} - E_{AF}$ and $\delta E_{AF} = E_{AF+ST} - E_{ST}$, are shown for total energy $E_{tot}$ and its two components hopping energy $E_t$ and interaction energy $E_U$, as a function of $U/t$.

## 3. Results and discussions

First, to grasp an overview of the two orders, let us consider magnetic behavior of $|\Psi_{AF+ST}\rangle$ and $|\Psi_{AF}\rangle$, and metal-insulator (Mott) transitions (MIT) in $|\Psi_{ST}\rangle$ and $|\Psi_{FS}\rangle$. Shown in Fig. 2(a) is the $\mathbf{q}=(\pi,\pi)$ element of spin structure factor $S(\mathbf{q}) = (1/N_s)\sum_{i,j}\langle \mathbf{S}_i \cdot \mathbf{S}_j \rangle\, e^{-i\mathbf{q}\cdot(\mathbf{r}_i-\mathbf{r}_j)}$ calculated with the optimized states. For $|\Psi_{AF+ST}\rangle$ and $|\Psi_{AF}\rangle$, as $U/t$ increases, $S(\pi,\pi)$ deviates from that of $|\Psi_{ST}\rangle$ and $|\Psi_{FS}\rangle$ and suddenly increases at $U_c$ and becomes proportional to the system size $L^2$ for $U > U_c^{AF+ST} \sim 5.0t$ and $U > U_c^{AF} \sim 1.5t$, respectively. This behavior of $S(\mathbf{q})$ and the fact that the optimized values of $\widetilde{\Delta}_{AF}$ becomes finite for $U > U_c$ indicate that AF long-range orders exist in the large-$U$ side of $U_c$. In addition, we confirm that an MIT simultaneously occurs at $U_c$, by monitoring the vanishing of Fermi surface in the momentum distribution function. This transitions are a Slater type rather than a Mott type, because (pure Mott-type) MIT's occur in $|\Psi_{ST}\rangle$ and $|\Psi_{FS}\rangle$ at much larger values: $U_c^{ST}/t \sim 9.0$ and $U_c^{FS}/t \sim 7.5$, where $S(\pi,\pi)$ exhibits a cusp, as shown in Fig. 2(a).

Next, we compare the optimized energies among the above four states, which are shown in Fig. 2(b) as a function of correlation strength. It is natural that $|\Psi_{AF+ST}\rangle$ is the lowest in energy. In a weakly correlated regime, $|\Psi_{FS}\rangle$ and $|\Psi_{AF}\rangle$ has high energies, because the exact ground state at $U=0$ has a Peierls phase ($\theta \neq 0$), namely, an appreciable current flows, but this phase cannot be appropriately cancelled in $|\Psi_{FS}\rangle$ and $|\Psi_{AF}\rangle$, where $\widetilde{\theta} = \phi = 0$. On the other hand, $|\Psi_{ST}\rangle$ and $|\Psi_{AF+ST}\rangle$ have a phase parameter $\widetilde{\theta}$, which is adjusted according to the Peierls phase $\theta$ in the Hamiltonian. In this regime of $U/t$, the effect of ST (AF) order is predominant (subordinate). For $U > U_c$, the two orders coexist and seems to cooperatively contribute toward reducing energy. In contrast, a ST state and a $d$-wave superconducting state are mutually exclusive [11].

We pursue the origin of the stability of $|\Psi_{AF+ST}\rangle$ more in detail. To this end, we first analyze the total energy $E_{tot}$ into kinetic part $E_t$ and interaction part $E_U$. Then, we estimate the quantities $\delta E_{ST} = E_{AF+ST} - E_{AF}$ and $\delta E_{AF} = E_{AF+ST} - E_{ST}$ for each component $E_{tot}$, $E_t$ and $E_U$. $\delta E_{ST}$ [$\delta E_{AF}$] indicates the contribution of the ST [AF] order. In Fig. 2(c), each elements are shown as function of $U/t$. We find from $\delta E_{ST}$ that the ST order develops by the gain in kinetic energy for any $U/t$, whereas from $\delta E_{AF}$, we find the source of stability in AF order is switched from the gain in the interaction energy for $U_c^{AF+ST} < U < U_c^{ST}$ to the kinetic energy for $U_c^{ST} < U$. $\delta E_{ST}^t$ and $\delta E_{ST}^U$ have cusps at $U_c^{AF}/t$ and $U_c^{AF+ST}/t$.

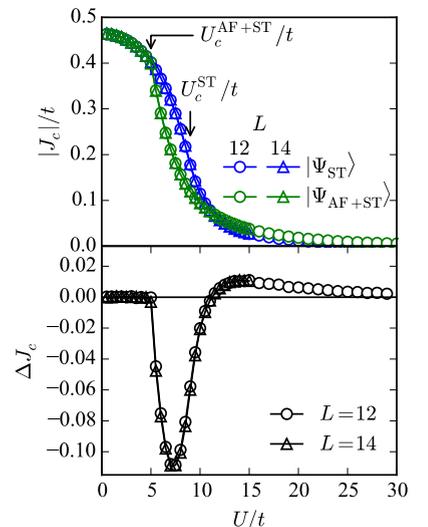

Fig. 3. (upper panel) Loop current $|J_c|$ is compared between $|\Psi_{AF+ST}\rangle$ and $|\Psi_{ST}\rangle$ as a function of $U/t$ for $L=12,14$ in the $\pi/2$-flux case. (lower panel) Difference of $|J_c|/t$ between the two states in the upper panel: $\Delta J_c = |J_c^{AF+ST}|/t - |J_c^{ST}|/t$.

We turn to the local loop current, which is calculated from,

$$J_c = \frac{it}{N}\sum_{i\in A}\langle \hat{a}_i^\dagger(\hat{b}_{i+x}+\hat{b}_{i-x})e^{i\theta} + \hat{a}_i^\dagger(\hat{b}_{i+y}+\hat{b}_{i-y})e^{-i\theta} - \text{H. c.}\rangle. \quad (6)$$

To consider the effect of AF order on $J_c$, we plot $J_c/t$ with respect to $|\Psi_{AF+ST}\rangle$ and $|\Psi_{ST}\rangle$ in the upper panel of Fig. 3. Both currents are monotonically decreasing function of $U/t$, and rapidly drop near the respective MIT points. In the lower panel, we show the difference of $|J_c|/t$: $\Delta J_c = |J_c^{AF+ST}|/t - |J_c^{ST}|/t$. $\Delta J_c$ once becomes negative, simply because $|\Psi_{AF+ST}\rangle$ becomes insulating at a smaller $U/t$ than $|\Psi_{ST}\rangle$. The sign of $\Delta J_c$ is reversed at $U/t \sim 11.0$, over which both states become insulating. The reason of positive $\Delta J_c$ in the insulating regime is as follows: The mobility of fermions is determined by $t/U$ [or $(t/U)^2$ for currents in square plaquette] if the spin configuration is antiparallel. If it is parallel, fermions cannot move owing to Pauli exclusion principle. Thus, Pauli principle disturbs currents. If there is an (no) AF order as in $|\Psi_{AF+ST}\rangle$ ($|\Psi_{ST}\rangle$), Pauli principle is less (more) effective. Thus, a current easily flows in $|\Psi_{AF+ST}\rangle$.

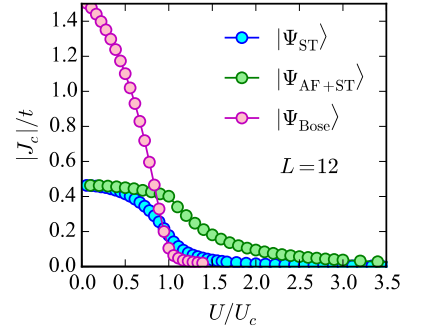

Fig. 4. Comparison of loop currents among a boson [7] and fermions for two wave functions for $L = 12$ in the $\pi/2$-flux case. Notice the difference of MIT points: $U_c^{ST} \sim 9.0t$, $U_c^{AF+ST} \sim 5.0t$, $U_c^{Bose} \sim 18.0t$.

Finally, we compare a feature of $J_c$ between fermions and bosons (Fig. 4). As mentioned, Pauli principle disturbs movements of fermions. Therefore, the mobility of bosons should be greater than that of fermions in equivalent conditions. Consequently, MIT points $U_c/t$ in bosons are much larger than those of fermions. Thus, in the insulating state, the value of $(t/U)^2$ is by far larger for bosons than for fermions. Such an account is reflected in the behavior of tails of $J_c$ in the insulating regime ($U > U_c$) in Fig. 4.

In this article, we focused on the fermionic-Hubbard model with $\pi/2$-staggered flux. We would like to extend this research to other magnitude of flux and doped cases.

**Acknowledgements**: A part of the numerical computations was carried out at the Cyber-science Center, Tohoku University. This work is partly supported by JSPS KAKENHI Grant Number 25287104.